\documentclass[
twocolumn,aps,preprintnumbers, amsmath,amssymb]{revtex4}

\usepackage{graphicx}
\usepackage{dcolumn}
\usepackage{bm}
\usepackage{comment}

\newcommand\ee{\end{equation}}
\newcommand\be{\begin{equation}}
\newcommand\eea{\end{eqnarray}}
\newcommand\bea{\begin{eqnarray}}
\newcommand{\sfrac}[2]{{\textstyle\frac{#1}{#2}}}
\newcommand\di{\partial}

\begin{document}

\preprint{INT-PUB-11-009}

\title{Hall viscosity from effective field theory}

\author{Alberto Nicolis}
\email{nicolis@phys.columbia.edu}

\affiliation{%
Physics Department and Institute for Strings, Cosmology, and 
Astroparticle Physics,\\
Columbia University, New York, NY 10027, USA
}%

\author{Dam Thanh Son}
\email{dtson@uw.edu}

\affiliation{%
Institute for Nuclear Theory, University of Washington, 
Seattle, Washington 98195-1550, USA
}%

\date{\today}

\begin{abstract}
For two-dimensional non-dissipative fluids with broken parity, we show
via effective field theory methods that the infrared dynamics
generically exhibit Hall viscosity---a conservative form of viscosity
compatible with two-dimensional isotropy.  The equality between the
Hall viscosity coefficient and the ground state's intrinsic angular
momentum density follows straightforwardly from their descending from
the same Lagrangian term of the low-energy effective action.  We show
that for such fluids sound waves are not purely longitudinal, but
acquire an elliptical polarization, with transverse-to-longitudinal
aspect ratio proportional to frequency.  Our analysis is fully
relativistic, thus providing a natural
description of (2+1) dimensional relativistic fluids with broken parity.

\end{abstract}

\maketitle

\noindent
{\em Introduction.}  At large distances and long times, the behavior
of any fluid can be described by the equations of fluid dynamics---the
continuity equation and the Navier-Stokes equation.  These equations are
in essense the equations of mass and momentum conservation,
\begin{align}
  \di_t\rho + \di_i (\rho v^i)      & =0\label{continuity}\\
  \di_t (\rho v^i) + \di_j T^{ij} & =0\label{NS}
\end{align}
coupled with an equation expressing the stress tensor $T^{ij}$ via
the $\rho$ and $v^i$.  The latter can be written down based on
symmetry considerations and involves shear and bulk viscosities.

In two dimensional fluids, it has been noticed some time
ago~\cite{ASZ,Avron-JSP} that new terms can be added into the
hydrodynamic equation if one relaxes the condition of parity and time
reversal invariance.  Namely, one can include into the stress tensor
the Hall viscosity, which is odd under these discrete symmetry,
and enters the hydrodynamic equations at the same order as the
conventional viscosities.  By nature, the Hall viscosity is
dissipationless.

The Hall viscosity has been investigated mostly in gapped systems,
where it is computed by subjecting the system under consideration to a
shear metric perturbation which is slowly changing with time.  The
Hall viscosity has been shown to be related to a Berry phase.  It was
found that the Hall viscosity is proportional to the density of
intrinsic angular momentum: $\eta_A= \frac12 \bar s \bar n$ where
$\bar n$ is the particle number density and $\bar s$ is the average
spin per particle (so $\bar s\bar n$ is the area density of intrinsic
angular momentum).

In this paper we construct a theory of a two-dimensional compressible
fluid with broken spatial parity.  We construct such a theory by
extending the action of a perfect fluid to include parity-breaking
effects.  As the theory is based on an action principle, it does not
contain dissipation effects.  We find that, when the leading
parity-breaking term is included into the Lagrangian, the equations of
motion describe a fluid with a nonzero intrinsic angular momentum
density.  In this fluid, in order to have a symmetric stress tensor,
in general the momentum density must be defined to contain a term
proportional to the derivatives of the intrinsic angular momentum
density, in addition to the $\rho {\vec v}$ term.

However, we find that in one particular case, namely when the
intrinsic angular momentum density is proportional to the particle
number density (i.e., the intrinsic angular momentum per particle is
constant), there is an alternative formulation of the hydrodynamic theory
where the momentum density remains $\rho {\vec v}$, but the stress
tensor now contains the Hall viscosity.  The value of the Hall
viscosity is exactly half of the intrinsic angular momentum
density.

Thus, in our compressible fluid, the Hall viscosity can be thought of
as an alternative description of a two-dimensional fluid with a
constant intrinsic angular momentum per particle.

\vspace{.3cm}

\noindent
{\em The setup.}  We will use the field-theoretical description of
fluids introduced in Ref.~\cite{DGNR}, which is based on previous work
on the field theory of solids~\cite{Soper:1976,Leutwyler:1996er} and
supersolids~\cite{Son:2005ak}.  We refer the reader to
Ref.~\cite{DGNR} for details (see also Refs.~\cite{NS,ENRW}). The
starting point is the parameterization of a fluid's configuration
space by giving at time $t$ the comoving (or ``Lagrangian'')
coordinates $\phi^I$ as functions of the volume elements' physical (or
``Eulerian'') positions $\vec x$:
\be
  \phi^I= \phi^I (\vec x ,t ) \;,  \qquad I=1, \dots , d \; , \label{phi}
\ee
where $d$ is the number of spatial dimensions.  This is
completely equivalent to the inverse point of view
where one gives the physical positions $\vec x$ as
functions of the comoving ones, $\vec x = \vec x (\phi^I, t)$ 
(this parametrization is adopted, e.g., 
in Ref.~\cite{JNPP}).  The advantage of using (\ref{phi})
is that $\phi^I$ can be treated as scalar fields in a field theory
with spacetime symmetries. We can thus construct local Lagrangians
with the desired symmetries following the usual rules of (effective)
field theory.

In a solid, the natural choice for the comoving coordinates $\phi^I$
is to point along the crystal axes.  For a fluid, there is an
arbitrariness in assigning comoving coordinates to the individual
volume elements.  If the fluid is at rest in the infinite past at
constant density $\rho_0$, then the most convenient choice is to
identify the comoving coordinates with the physical ones at
$t\to -\infty$
\be \label{choice}
\phi^I = x^I \;  .
\ee
With this choice of coordinates, the internal symmetries the dynamics 
must obey are \cite{DGNR}
\bea
\phi^I & \to & \phi^I + a^I \; ,  \qquad I=1, \dots , d  \label{shift}\\
\phi^I & \to & R^I {}_J \, \phi^J  \; , \qquad  R \in SO(d)   \label{rotate} \\
\phi^I & \to & \xi^I (\phi) \;, \qquad \det \frac{\di \xi^I}{\di \phi^J} = 1
    \label{diff}
\eea
Equations.~(\ref{shift}) and (\ref{rotate}) correspond to the physical
homogeneity and isotropy of the fluid's internal
space. Eq.~(\ref{diff}) is what distinguishes a fluid from an
isotropic solid: displacing volume elements without compressing or
dilating the fluid anywhere does not cost any energy.

Our main goal will be to construct parity-breaking hydrodynamics for
non-relativistic fluids, but we find it convenient to keep our
analysis fully relativistic and to take the non-relativistic limit
only when needed.  Our symmetries are therefore
eqs.~(\ref{shift}--\ref{diff}) plus the $(d+1)$-dimensional Poincar\'e
group, under which the $\phi^I$'s behave as scalars.  We will use the
$(-,+, \dots, +)$ signature for the metric.  A fundamental object that
we will use extensively is
\begin{equation}
\begin{split}
J^\mu & \equiv  \epsilon^{\mu\alpha_1 \dots \alpha_d} \, 
  \di_{\alpha_1} \phi^1 \dots \di_{\alpha_d} \phi^d \\
& = \frac{1}{d!} \epsilon^{\mu\alpha_1 \dots \alpha_d} \, 
  \epsilon^{I_1 \dots I_d} 
\, \di_{\alpha_1} \phi^{I_1} \dots \di_{\alpha_d} \phi^{I_d} \, . \label{J}
\end{split}
\end{equation}
(We define the $(d+1)$-dimensional $\epsilon$ tensor by
$\epsilon^{01\dots d} = +1$.)  It is a vector under Poincar\'e, and is
invariant under our internal symmetries (\ref{shift}--\ref{diff}). We
will denote its norm by $b$,
\be
b \equiv \sqrt{-J_\mu J^\mu} = \sqrt{\det B^{I\!J}} \, , 
  \qquad B^{I\!J} \equiv \di_\mu \phi^I \di^\mu \phi^J  , \label{b}
\ee
which gives a measure of the compression level of the fluid.  The
velocity field $u^\mu$ of a fluid configuration is defined as a unit
timelike vector aligned with $J^\mu$
\be
J^\mu = b \, u^\mu \, . \label{u}
\ee
The current $J^\mu$ is identically conserved:
\be
\di_\mu J^\mu = 0  \quad \mbox{(identically)}. \label{dJ}
\ee
Equations~(\ref{u}), (\ref{dJ}) invite the interpretation of $b$ as the
number density of fluid elements, or of fluid points, whose
conservation law should indeed be an identity.

\vspace{.3cm}

\noindent
{\em The Lagrangian and the stress-energy tensor.}  According to
standard effective field theory logic, at low energies and momenta we
should organize the dynamics as an expansion in derivatives.  The
low-energy Lagrangian thus takes the form \cite{DGNR}
\be
{\cal L} = F(b) + \mbox{higher derivatives} \, ,
\ee
where $F$ is a generic function, to be determined from the equation of
state. More precisely, from the lowest derivative part of the
Lagrangian, $F(b)$, one gets the standard perfect fluid stress energy
tensor,
\be
T_0^{\alpha\beta} = (\rho+p) \, u^\alpha u^\beta 
  + p \, \eta^{\alpha\beta} \, , \label{T0}
\ee
with energy density and pressure given by \cite{DGNR}
\be
\rho = - F(b) \;, \qquad p = F(b) -F'(b) b \, . \label{rhop}
\ee
For a barotropic fluid the equation of state is an algebraic relation
between $p$ and $\rho$: $p = p(\rho)$. This translates, via
Eq.~(\ref{rhop}), into a differential equation for $F$.

The higher derivative terms generically involve at least two more
derivatives, on top of those already present in objects like $J^\mu$
and $b$, which involve one derivative per field.  Indeed, the shift
symmetry (\ref{shift}) forces each field $\phi^I$ to be acted upon by
at least one derivative.  So, at lowest order in derivatives the only
object that is invariant under the internal symmetries and that
transforms covariantly under Poincar\'e is $J^\mu$ itself, along with
functions thereof like $b$.  If we add only one more derivative, the
general structure we expect is a generic function of $b$, times a
derivative on $J^\mu$, times extra powers of $J^\mu$,
\be
\Delta {\cal L} = f(b) \, \di J \, J^n   \; ,
\ee
with suitable contractions of the Lorentz indices. Notice that a
derivative acting on a function of $J^\mu$ is still rewritable in the
above form, via the chain rule. The derivative cannot be contracted
with the $J$ it acts on, because of (\ref{dJ}). Moreover, all
contractions between two $J$'s both belonging to the $J^n$ piece just
redefine $f(b)$, since $J^\mu J_\mu = -b^2$. Finally, the $J^n$ piece
cannot have more than one $J$ contracted with an $\epsilon$ tensor,
because of symmetry. So, for generic dimensionality $d$, at the
one-derivative level the only possibility is
\begin{equation}
\begin{split}
\Delta {\cal L} = f(b) \, \di_\mu J_\nu \, J^\nu J^\mu  
  & = -\sfrac12 f(b) \, \di_\mu b^2 \, J^\mu \\
 & =  \di_\mu g(b) \, J^\mu   \,, 
\end{split}
\end{equation}
which vanishes upon integrating by parts \footnote{The function $g$ is
defined as $g(b) = - \int f(b) \, b \, db$.}. One is thus led to
consider terms with more derivatives.

However in $2+1$ dimensions we have one  more possibility:
\be
\Delta {\cal L} = f(b) \, \epsilon^{\mu\nu\rho} J_\mu \, \di_\nu J_\rho \, ,
   \qquad d =2 \, , \label{DeltaL}
\ee 
This is invariant under all our symmetries, and should generically be
there in two-dimensional fluids where parity is broken.  Its
contribution to the stress tensor will involve a single derivative
acting on physical quantities, like the velocity field for
instance. In this sense it is of the same order as viscosity---it
appears at the same order in the derivative expansion. However, coming
from a Lagrangian term, this will be a {\em conservative} (i.e.,
non-dissipative) form of viscosity.  Our claim, which we are now going
to prove, is that (\ref{DeltaL}) is the Lagrangian description of Hall
viscosity.

To compute the stress-energy tensor we vary $\Delta {\cal L}$ with
respect to the metric.
After a straightforward computation we get
\begin{multline}
\Delta T^{\alpha \beta}  \equiv   2  \, \frac{\delta (\Delta S)}{\delta g_{\alpha \beta }} =
-  \frac{d \log f}{d \log b }\,  \Delta {\cal L} \, (\eta^{\alpha\beta} + u^\alpha u^\beta)  \\
+  4  f \, \epsilon^{ \nu \mu (\alpha} J^{\beta)} \di_\nu J_\mu 
+  2 \di_\nu f \, \epsilon^{\nu \mu (\alpha } J^{\beta)}  J_\mu
\end{multline}
where we made use of the identity
\be
B^{-1}_{I\!J} \, \di^\alpha \phi^I \di^\beta \phi^J 
  = \eta^{\alpha\beta} + u^\alpha u^\beta \, .
\ee
At the one-derivative level, the full stress energy tensor is
\be
T^{\alpha\beta} = T_0^{\alpha\beta} + \Delta T^{\alpha \beta}  \; ,
\ee
where $T_0^{\alpha\beta}$ is the perfect fluid part, eq.~(\ref{T0}).

\vspace{.3cm}

\noindent
{\em Intrinsic angular momentum and Hall viscosity.}  A crucial fact
is that the otherwise arbitrary coefficient $f(b)$ turns out to be
related to the angular momentum density for static configurations. The
total angular momentum is defined as
\be 
L = \epsilon^{ij}\! \int\! d^2 x \, x^i T^{0j}
\ee
(it is a two-dimensional scalar.)
For vanishing fluid velocity, $T^{0i}$ reduces to
\be
T^{0i} = \Delta T^{0i} = - \epsilon^{ij} \, \di_j (f b^2) \, .
\ee
Upon plugging this into the expression for $L$, integrating by parts,
and discarding the boundary term which vanishes for a finite-size
fluid (if we take the boundary outside the fluid), we get an angular
momentum surface density
\be
\ell \equiv \frac{d L}{d S} = - 2 \, f b^2 \,.  \label{ell}
\ee
The fact that the angular momentum density does not vanish whhen the
fluid is at rest is not surprising---parity breaking allows it.  This
situation occurs, for example, in the A-phase of superfluid
helium-3~\cite{Vollhardt}.

We now take the nonrelativistic limit.  In this limit, $b$ is the mass
density $\rho$, so we will use $b$ and $\rho$ interchangeably~\footnote{Most
of the equations derived below remain valid for a fluid moving slowly, 
but with a relativistic equation of state; one only needs to replace
$\rho$ by $\rho+p$ in all formulas.}.
At lowest order in the fluid velocity, our corrections read
\begin{align}
\Delta T^{00} & =  {\cal O}(\di \ell v) \\
\Delta T^{0i} & =  \sfrac12 \epsilon^{ij} \, \di_j \ell 
  +  {\cal O}(\di \ell v^2) \\
\Delta T^{ij} & =  -\big(\ell - \sfrac12 b \, \ell'  \big) \, 
  (\epsilon^{kl} \di_k v_l) \,  \delta^{ij} \nonumber\\
&\quad + \sfrac12 \big( \epsilon^{ik} v^j \di_k \ell 
  + i \leftrightarrow j \big)  +{\cal O}(\di \ell v^3) \, ,\label{DTij}
\end{align}
where we parameterized everything in terms of the angular momentum
density $\ell(b) = -f(b) b^2$, $\ell'$ stands for the derivative of
$\ell$ with respect to $b$, and the $\di$ inside the ${\cal
O}(\dots)$'s denotes schematically a spatial gradient.
One can check that all ${\cal O}(\dots)$ are suppressed in the
nonrelativistic limit and can be dropped.

First, we see that $\Delta T^{ij}$ does not resemble the stress tensor
associated with Hall viscosity.  Its traceless part contains terms
proportional to the gradient of the angular momentum density $\ell$,
but not gradients of the velocity.
Moreover, the non-relativistic momentum density is not the naive $\rho
v^i$, being corrected by a term proportional to the gradient of the
angular momentum,
\be
T^{0i} = \rho v^i + \sfrac12 \epsilon^{ij} \, \di_j \ell \; .
\ee
That the momentum density involves the gradient of the angular
momentum density is nothing new~\cite{Martin:1972,Vollhardt}.
Galilean invariance then dictates that the stress tensor must contains
terms of the form $v\di\ell$, as in Eq.~(\ref{DTij}) (in contrast, the
term proportional to $\epsilon^{kl}\di_k v_l$ in Eq.~(\ref{DTij})
is not dictated by Galilean invariance).

The extra term in the
momentum density does not contribute to the total momentum
$\int\!d^2x\, T^{0i}$, and has zero divergence: $\di_i\Delta
T^{0i}=0$.  For this reason, the continuity equation
remains~(\ref{continuity}), and the momentum conservation can still be
written as Eq.~(\ref{NS}), but with a new stress tensor,
\begin{equation}
  \di_t(\rho v^i) + \di_i (T^{ij} + \Sigma^{ij}) = 0, \quad
    \Sigma^{ij} =  -\sfrac12 \epsilon^{ij} \, \ell' \, \di_k (b \, v^k) 
\end{equation}
$\Sigma^{ij}$ has been defined so that $\di_i \Delta T^{0i}=\di_j
\Sigma^{ij}$.  The modified stress tensor is not symmetric.  It should be
expected since the symmetry of this stress tensor would lead to the
conservation of the naive angular momentum $\int\!d{\vec
x}\,\rho\,\epsilon^{ij}x^iv^j$, but we know that the conserved angular
momentum is actually the sum of the naive angular momentum and the
``spin,''
\begin{equation}
  \int\!d{\vec x}\, (\rho \,\epsilon^{ij}x^i v^j + \ell)
\end{equation}

In one particular case, however, we should be able to symmetrize
$\Sigma^{ij}$ via the Belinfante trick: when the angular momentum
density is proportional to the particle number density:
\be \label{assumption}
\ell = \lambda b \, , \qquad \lambda = {\rm const}  \; .
\ee
In this case the total ``spin'' is proportional to the total particle
number and is conserved by itself.  This is a non-trivial assumption,
which is obeyed for instance by fluids where the bulk of the angular
momentum is carried by individual spins or by bound states, so that in
first approximation it scales linearly with the number of particles in
the system. Under this assumption we can symmetrize $\Sigma^{ij}$ via
the addition of the total divergence of an antisymmetric tensor,
$\di_k \Lambda^{i[jk]} $, which does not affect momentum
conservation. We have to choose
\be
\Lambda^{i[jk]} = \sfrac1 2 \lambda b\big[ \epsilon^{ij} \,  v^k - \epsilon^{ik} \,  v^j - \epsilon^{jk} \,  v^i \big] \, ,
\ee
from which we get
\be
\Sigma^{ij} + \di_k \Lambda^{i[jk]} = - \sfrac1 2 \lambda \big[ \epsilon^{ik} \di_k (b \, v^j) +  i \leftrightarrow j \big] \, ,
\ee
which is symmetric, as desired. Putting everything together we get
\begin{align}
\tilde T^{00} & =  \rho \,, \\
\tilde T^{0i} & =  \rho v^i \,, \\
\tilde T^{ij} & =  p  \, \delta^{ij} + \rho v^i v^j + \Delta \tilde T^{ij} 
\end{align}
with
\begin{equation}
\begin{split}
\Delta \tilde T^{ij} & \equiv   \Delta T^{ij} + \Sigma^{ij} 
  + \di_k \Lambda^{i[jk]} \\
& =  - \sfrac 12 \lambda b \, \big[ (\epsilon^{kl} \di_k v_l) \, \delta^{ij} 
  +  \big( \epsilon^{ik} \di_k v^j 
  +  i \leftrightarrow j  \big) \big]  . \label{final_Tij}
\end{split}
\end{equation}
Notice that all derivatives of the number density $b$ canceled out, leaving 
us with derivatives of the velocity field only.
The form of $\Delta \tilde T^{ij} $ matches precisely  Hall 
viscosity~\cite{ASZ,Avron-JSP}.
It can be rewritten as
\be
  \Delta  \tilde T_{ij} = 
  - \eta_{\rm H} 
  ( \epsilon_{ik}\delta_{jl} + \epsilon_{jk}\delta_{il}) V_{kl} \, ,
 \quad V_{kl} = \tfrac12 (\di_k v_l + \di_l v_k)
\ee
with
\be
  \eta_{\rm H}  =  \sfrac 12 \lambda b =  \sfrac12 \ell \; .
\ee
The Hall viscosity coefficient is thus half the angular momentum
density.  This relationship between the Hall viscosity and the angular
momentum density has been shown to occur in gapped system using adiabatic
arguments~\cite{Read}.  In
our description of a compressible fluid, it emerges as 
a straightforward consequence of the
simplicity of the low-energy effective action. At next-to-lowest order
in the derivative expansion, all observables derive from a single
Lagrangian term, eq.~(\ref{DeltaL}). Moreover, our Lagrangian provides
a relativistic generalization of Hall viscosity.

\vspace{.3cm}

\noindent
{\em Sound wave propagation.}  We now investigate sound-wave propagation
in a compressible fluid with broken parity.  Sound waves are obtained 
from our field theoretical description by expanding
the Lagrangian at quadratic order in small perturbations. Explicitly,
if we consider small deviations from the static, homogeneous
configuration (\ref{choice}),
\be
\phi^a = x^a + \pi^a \; ,
\ee
where $\pi^a$ is (up to a sign) the displacement, 
the leading-order Lagrangian, to second order in the phonon
field $\vec \pi$, becomes \cite{DGNR,ENRW}
\be \label{quadratic}
F(b) \to w_0 \big[ \sfrac12 \dot {\vec \pi}^2  - \sfrac12  c_s^2 (\vec \nabla \cdot \vec \pi)^2 \big]   \; ,
\ee
where $w_0 \equiv -F'(1) = (\rho + p)_{b=1}$ is the background
enthalpy density (which becomes the mass density in the
nonrelativistic limit), and $c_s^2 \equiv
F''(1)/F'(1)=(dp/d\rho)_{b=1}$ is the sound speed.  As expected, only
the longtidinal part of $\vec \pi$ propagates.  The transverse modes
lack gradient energy---this is a direct consequence of our symmetry
(\ref{diff})---and do not propagate.

We now restrict to $d=2$ case, add our one-derivative correction,
Eq.~(\ref{DeltaL}), and expand it at second order in $\vec \pi$. By
using eq.~(\ref{u}) and
\begin{align}
u^0 & = 1+{\cal O}(\pi^2) \, , \quad  
  \vec u = - \dot {\vec \pi} + (\dot {\vec\pi}\cdot\nabla){\vec \pi}+ 
  {\cal O}(\pi^3)  \, , \\
b & =  1 + \vec \nabla \cdot \vec \pi + {\cal O}(\pi^2) \, ,
\end{align}
we get
\be
\Delta {\cal L} \to \sfrac12 \big[ \ell \, \epsilon^{ij} \dot \pi^i \ddot \pi^j - \ell' \, \di_k \pi^k \, \epsilon^{ij} \di_i \dot\pi^j  \big],
\ee
where $\ell$ and $\ell'$ are evaluated at $b=1$, we used the universal
relation between $f$ and $\ell$, eq.~(\ref{ell}), but we did {\em not}
assume Eq.~(\ref{assumption}), nor did we take the non-relativistic
limit. This correction to the quadratic Lagrangian (\ref{quadratic})
induces a mixing between longitudinal and transverse modes, as can be
immediately seen by decomposing the phonon field as $\vec \pi = \vec
\pi_L + \vec \pi _T$. To diagonalize the Lagrangian, one can introduce
a new phonon field $\vec \pi '$, with longitudinal and transverse
components $\vec \pi'_L$ and $\vec \pi'_T$, respectively, such that
\begin{align}
\pi^i_L & =  \pi'_L {}^i -\sfrac{\ell'}{2 w_0 c_s^2} \epsilon^{ij}  \dot \pi'_T {}^j \\
\pi^i_T & =  \pi'_T {}^i + \sfrac{1}{w_0} \Big[ \sfrac{\ell'}{2 c_s^2} - \ell \Big] \epsilon^{ij}  \dot \pi'_L {}^j \; .
\end{align}
One gets simply
\be
{\cal L} \to w_0 \big[ \sfrac12  \dot {\vec \pi} ' \, ^2  - \sfrac12  c_s^2 (\vec \nabla \cdot \vec \pi')^2 \big]  
+ \mbox{higher derivatives} \; ,
\ee
where now the higher derivative terms involve at least {\em two} more
derivatives, i.e.~they are on an equal footing with higher-order
corrections that we have been neglecting all along. It is thus
consistent to discard them---and in fact it would be inconsistent not
to.  We see that at the order we are working the spectrum of
perturbations is unaltered: $\vec \pi '_T$ does not propagate while
$\vec \pi '_L$ features a linear dispersion relation with propagation
speed $c_s$. However what is interesting is that this propagating mode
does not look purely longitudinal once expressed in terms of the
original field $\vec \pi$: it is {\em elliptically} polarized,
alternating between the longitudinal and the transverse direction,
with an aspect ratio proportional to frequency,
\be
\frac{\pi_T}{\pi _L } \simeq  \frac{\omega}{w_0} \Big[ \frac{\ell'}{2 c_s^2} -\ell \Big] \; .
\ee
We do not expect generically a cancellation between the two terms
inside the bracket: one depends on the sound speed while the other
does not, and, in particular, for a non-relativistic fluid the latter
will be negligible with respect to the former.

\vspace{.3cm}

\noindent
{\em Conclusions.}  In this paper we have considered a field theory
description of a (2+1) dimensional fluid with broken parity.  We have shown
that the fluid naturally has a finite density of angular momentum, and
that in the nonrelativistic limit it allows a description involving Hall
viscosity if the intrinsic angular momentum per particle is constant.
The Hall viscosity is shown to be exactly half of the area density of
the intrinsic angular momentum.  We also investigated the propagation
of sound wave in such a fluid.  While we did not find modifications to
the sound speed, we found that the sound wave is a superposition of
transverse and longitudinal waves.

Although in this paper we have taken the point of view that the
parity-breaking effect is the next-to-leading order effect in the
derivative expansion, one can regard the Lagrangian ${\cal L} +
\Delta{\cal L}$, without higher order terms, as a complete description of
a hypothetical fluid, even when the $\Delta{\cal L}$ is of the same
order as ${\cal L}$.  For example, one can consider a fluid of
particles with very high spin.  The hydrodynamic equations that we
derived do not suffer inconsistencies when truncated to the order that
we have considered.

\vspace{.3cm}
 
\noindent {\em Acknowledgements.}  AN would like to thank Walter
Goldberger and Omid Saremi for useful discussions.  DTS thanks Nick
Read for discussions and valuable comments.  The work of AN is
supported in part by the DOE (DE-FG02-92-ER40699).

\end{document}